\documentstyle [11pt,epsf]{article}
\begin{document}
%
%
\newcommand{\Abs}[1]{|#1|}
\newcommand{\EqRef}[1]{(\ref{eqn:#1})}
\newcommand{\FigRef}[1]{fig.~\ref{fig:#1}}
\newcommand{\Abstract}[1]{\small
   \begin{quote}
      \noindent
      {\bf Abstract - }{#1}
   \end{quote}
    }
\newcommand{\FigCap}[2]{
\ \\
   \noindent
   Figure~#1:~#2
\\
   }
%
%
%
%
\title{Lyapunov exponents and anomalous diffusion of a Lorentz gas with
infinite horizon using approximate zeta functions}
\author{Per Dahlqvist \\
Mechanics Department \\
Royal Institute of Technology, S-100 44 Stockholm, Sweden\\[0.5cm]
}
\date{}
\maketitle

\ \\ \ \\

%
\Abstract{We compute the Lyapunov exponent, generalized Lyapunov exponents and
the diffusion constant for a Lorentz gas on a square lattice, thus having
infinite horizon. Approximate zeta functions, written in terms of probabilities
rather than periodic orbits, are used in order to avoid the convergence
problems
of cycle expansions. The emphasis is on the relation between the analytic
structure of the zeta function, where a branch cut plays an important role, and
the asymptotic dynamics of the system. We find a diverging diffusion constant
$D(t) \sim \log t$ and a phase transition for the generalized Lyapunov
exponents.

%
}

\ \\
\section{Introduction}

Maybe the most well known measure of a chaotic system is the
Lyapunov exponent. In the theory of chaotic dynamics one is of course
interested in calculating this and similar quantities, either by finding
analytical estimates or by devising effective calculation schemes but often one
find oneself compelled to numerical simulation. This is unsatisfactory
since it is not an easy task to extract information on the asymptotic
behaviour from numerical data.

Various
averages of chaotic systems are obtainable via transfer operators and
and their Fredholm determinants or
{\em zeta functions}. This is a beautiful formalism but the best results are
obtained for a very restricted class of chaotic systems, namely those
fulfilling
Axiom-A. This is because Axiom-A guarantees nice
analytical features of the zeta functions \cite{Polli,Rue1,Rue2,Rugh} which
enables fast convergent cycle expansions to deduce its leading zeros
\cite{AAC}.
Applications of cycle expansions to non Axiom-A systems are not very
successful \cite{TS,PDreson}.

In this paper we will study a system which is far from the
textbook 1-d Axiom-A map, namely the two-dimensional Lorentz gas on a square
lattice. This is a Hamiltonian
system with two degrees of freedom, continous time
and with an infinite symbolic dynamics.
We will demonstrate that zeta functions
may be of great use even here. The key point is that we will avoid writing the
zeta function in terms of periodic orbits as that would lead us to divergence
problems that we could not handle. The price we will pay is that our zeta
functions are no longer exact, but they are approximate in a sense that won't
effect the leading zero very much. The averages we will compute are directly
related to the motion of this leading zero with respect to  variations of a
parameter. However,
matters will be complicated if there are non-analyticities in the vicinity
of the leading zero, like branch cuts. This is, we believe,
a generic feature of
non Axiom-A systems. Such singularities will
cause problems for cycle expansions but do
carry important information about the asymptotic dynamics.
In the Lorentz gas there will be a branch cut
connected to the existence of the infinite horizon. It will not prevent the
Lyapunov exponent from being well defined but will yield a diverging diffusion
constant. It will also imply a phase transition for the generalized Lyapunov
exponents.

In section 2 we review the necessary theory. In section 3 we perform all
calculations and we then end with some comments
in section 4.

\section{Theory}

\subsection{Lyapunov exponents and zeta functions}

The largest Lyapunov exponent is defined by
\begin{equation}
\lambda = \lim_{t \rightarrow \infty} \frac{1}{t} \log | \Lambda(x_0,t)|  \ \ ,
\end{equation}
provided the limit exists and is independent of the initial point $x_0$ (except
for a set of measure zero: namely the periodic points).
$\Lambda(x_0,t)$ is the largest eigenvalue of the
Jacobian along the trajectory starting at $x_0$ and evolving during time $t$.
Using ergodicity this may be rewritten as a phase space average
\begin{equation}
\lambda =  \frac{1}{t}
\int \mu(dx) \log | \Lambda(x_0,t)|\equiv
\frac{1}{t}< \log | \Lambda(x_0,t)|>
\   \   , \end{equation}
where $\mu(dx)$ is the invariant density.

We will now formulate the Lyapunov exponent in terms of evolution operators
and zeta functions.
Consider the following evolution operator acting on a phase space density
$\Phi(x)$ according to
\begin{equation}
{\cal L}_w^t  \Phi(x)=\int w(x,t)\delta (x-f^t(y))\Phi(y)dy  \label{eqn:evdef}
\ \ .
\end{equation}
It is essential for the further development that
the weight $w(x,t)$ is multiplicative along the flow.
We intend to evaluate the trace of this operator as that gives the
expectation values of the weight (see appendix)
\begin{equation}
<w>=\lim_{t \rightarrow \infty} tr {\cal L}_w^t  \label{eqn:start} \   \   .
\end{equation}

The trace naturally turns out to be a sum over the periodic orbits
\begin{equation}
tr {\cal L}_w^t  = \int w(x,t) \delta (x-f^t(x))dx=
\sum_p T_p \sum_{n=1}^{\infty} w_p^n \frac{\delta(t-nT_p)}
{\Abs{det(1-M_p^n)}}
 \ \ ,
\label{eqn:tracedef}
\end{equation}
where $n$ is the number of repetitions of primitive orbit $p$, having period
$T_{p}$, and  $M_{p}$ is the Jacobian (transverse to the flow). $w_p$ is the
weight associated with cycle $p$.

Zeta functions are introduced by observing that the trace may we written as
\begin{equation}
tr {\cal L}_w^t = \frac{1}{2\pi i}
\int_{-\infty-ia}^{\infty-ia} e^{ikt}\frac{Z_w'(k)}{Z_w(k)}dk   \ \ .
\label{eqn:tracefour}
\end{equation}
For a Hamiltonian system with two degrees of freedom the zeta function reads
\cite{flows}
\begin{equation}
     Z_w(k)=\prod_{p}\prod_{m=0}^{\infty}
          \left(1-w_p \frac{e^{-ikT_{p}}}
    {\Abs{\Lambda_{p}} \Lambda_{p}^{m}}\right)^{m+1}
          \ \ ,                       \label{eqn:Zw}
\end{equation}
where $\Lambda_{p}$ is the expanding
eigenvalue of $M_p$.

Such infinite products over periodic orbits are often troublesome because they
diverge beyond the  first (nontrivial) zero. It is essential that the constant
$a$ in eq.  \EqRef{tracefour} is
sufficiently large so that the products converge.
In our subsequent calculation it suffices if $a$ is small and positive.

Cycle expansions of the zeta functions generally have better convergence
properties since they
converge up to the first singularity \cite{AAC}.
The zeta function is entire for Axiom-A system which makes cycle expansions
very successful for this special case.
In this paper we will consider cases where the leading zero is also a
singularity (branch point) so a cycle expansion will not converge even there
and the zeta function is rather useless as it stands. We return to these
problems in section 2.3.

We must now find a weight $w$ appropriate for computing the Lyapunov exponent.
The quantity whose average we are going to study is
$log |\Lambda(x_0,t) |$ which is certainly not multiplicative.
In one dimension we can study the average of the multiplicative
weight $w=|\Lambda(x_0,t) |^{\tau} $
and obtain the Lyapunov exponent by differentiation
\begin{equation}
\lambda= \lim_{t \rightarrow \infty} \frac{1}{t} \frac{d \; tr {\cal L}_\tau^t}
{d \tau } |_{\tau=0}    \  \  . \label{eqn:liatrace}
\end{equation}

The problem
to find the appropriate weighted operator in more dimensions is nontrivial and
the problem was recently solved \cite{Gabor}.
But for our purposes it will suffice to insert
the weight $w$ above into the operator ${\cal L}_w^t$ just defined.
The modification in \cite{Gabor}
to make $w$ exactly multiplicative is complicated but
minor and we don<t expect it to affect the leading
zero of the zeta function, and its vicinity, in which we are interested.

The leading zero $k_0(\tau)=-i\cdot h(\tau )$ is always on the
negative imaginary
axis, if $\tau>0$. It provides the leading asymptotic behaviour of the trace
provided that there is a gap until next zero or singularity.
Then we have
\begin{equation}
\lambda= \frac{d}{d\tau} h(\tau)  \  \   .
\end{equation}
Generalized Lyapunov exponents \cite{Beck} $\lambda (\tau )$
are defined by considering the
scaling behaviour of \\
$<|\Lambda(x_0,t)|^\tau>$:
\begin{equation}
<|\Lambda(x_0,t)|^\tau> = \lim_{t \rightarrow \infty}tr {\cal L}_\tau^t
= e^{\lambda(\tau)\tau t}   \  \  ,
\end{equation}
so that
\begin{equation}
\lambda(\tau ) = \frac{h(\tau)}{\tau}   \   \  ,
\end{equation}
provided again that the leading zero is isolated.
The ordinary Lyapunov exponent is recognized as the limit
$\lambda= \lim_{\tau \rightarrow 0} \lambda(\tau)$.

\subsection{Diffusion coefficients and zeta functions}

We will consider the Lorentz gas obtained by unfolding the Sinai billiard.
The coordinate in the unfolded system is called $\hat{x}$. The corresponding
vector in the billiard (or the {\em unit cell}) is $x$. They are related by
translation $\hat{x}-x \in T$ where $T$ is the group of translations
building up the Lorentz gas from the unit cell.

The diffusive properties can be extracted from the average
\begin{equation}
< e^{\beta \cdot (\hat{f}^t(x_0)-x_0)} >_{x_0} \ \  .
\end{equation}
The average is taken over one unit cell.
Again we must perform the trick to introduce a multiplicative weight and then
by
differentiation extract the average in which we are interested

It was demonstrated in ref \cite{CEG} that this average can be computed
by considering the dynamics in the unit cell only. This is obtained by
inserting the weight
\begin{equation}
w(x,t)=e^{\beta \cdot (\hat{f}^t(x)-x)}
\label{eqn:diffw}
\end{equation}
into the evolution operator \EqRef{evdef}.
The diffusion constant is now given by
\begin{equation}
D=\lim_{t \rightarrow \infty} \frac{1}{\nu t} \sum_{i=1}^{\nu}
< (\hat{f}(x_0)-x_0)^2>_{x_0}=
\lim_{t \rightarrow \infty} \frac{1}{\nu t}
\sum_{i=1}^{\nu} \frac{\partial^2}{\partial \beta_i^2}tr {\cal L}_\beta^t  \ \
{}.
\end{equation}
where the sum is taken over spatial components of the $2\nu$
dimensional phase  space.

\subsection{Approximate zeta functions}

In some recent publications we have investigated a way of approximating zeta
functions for intermittent systems \cite{PDreson,PDsin,PDLA}.
We call it the BER-approximation after the authors of ref \cite{BER}.
In an intermittent system laminar intervals are interrupted
by chaotic outbursts. Let $\Delta_i$ be the time elapsed between two
successive entries
into the laminar phase. The index $i$ labels the i'th
interval.
Provided the chaotic phase is {\em chaotic enough},
the lengths of the intervals $\Delta_i$
are presumed uncorrelated,
and $\Delta$ may be considered as a stochastic variable
with probability distribution $p(\Delta)$.
The zeta functions (unit weight $w=1$) may then be expressed in terms
of the Fourier transform of $p(\Delta)$
\begin{equation}
Z(k)\approx \hat{Z}(k)
\equiv 1-\int_{0}^{\infty}e^{-ik\Delta}p(\Delta)d\Delta   \   \ .
\label{eqn:ZBER}
\end{equation}
Due to the normalization of $p(\Delta )$ we see that leading zero $k_0=0$ is
by construction exact because of probability conservation.

In order to compute the
probability distribution we introduce
a surface of section (SOS).
This should, according to the BER prescription be put
on the border
between the laminar and chaotic phase.
We call the phase space of the SOS $\Omega$ and its coordinates $x_s$.
The  flight time to the next intersection is then a function of $x_s$:
$\Delta_s(x_s)$. The probability distribution then reads
\begin{equation}
p(\Delta)=\int_{\Omega} \delta (\Delta-\Delta_s(x_s)) \mu(dx_s) \ \ ,
\end{equation}
where $\mu(dx_s)$ is the invariant density, which is uniform
$\mu(dx_s)=dx_s/\int_\Omega dx_s$
for Hamiltonian ergodic system, assuming of course that the SOS coordinates
are canonically conjugate.

It is straightforward to include weights in this formalism, like
e.g. $w=|\Lambda(x_0) |^{\tau}$.
The local expansion factor $\Lambda_s(x_s)$ is also a function of $x_s$.
The zeta function
is then related to a generalized distribution
\begin{equation}
p_\tau(\Delta)=\int_{\Omega} |\Lambda_s(x_s)|^{\tau}
\delta (\Delta-\Delta_s(x_s)) dx_s     \ \ . \label{eqn:psos2}
\end{equation}
The zeta function $\hat{Z}_\tau (k)$ is obtained by inserting $p_\tau(\Delta)$
into \EqRef{ZBER}.

\section{Application to the Lorentz gas}

We now begin our study of the Lorentz gas on a square lattice. The lattice
spacing is unity, the disk circular with radius $R<1/2$, and the point particle
bouncing around in this array has unit velocity.
It is the unit cell of this system, the Sinai billiard \cite{Sin} whose
dynamics
we will study and this is indeed an intermittent system; there exist periodic
orbits with arbitrary small Lyapunov exponents $\log \Lambda_p/T_p$.
The disk will define our SOS. We use the two angles $\phi$ and $\alpha$ defined
in fig 1a  as coordinates.
The normalized measure is then $dx_s=d\phi \; d(\sin \alpha)/4\pi$.
Consider now a segment of the trajectory between two disk collisions.
This segment can be labeled according to the disk that would be hit in the
unfolded system, the label ${\bf q}=(n_x,n_y)$ is the associated lattice vector
\cite{PDsin}. It is easy to realize that only disks
associated with coprime lattice vector may
occur.

\begin{figure}[p]
\vspace{15cm}
\caption{
a) The Sinai billiard with definitions
of the variables $\phi$ and $\alpha$.
b) The unfolded system with free directions (corridors) indicated.
c) The region of integration.}
\end{figure}

The purpose is now to apply the BER approximation to this system and compute
Lyapunov exponents and diffusion constants, cf refs \cite{PDsin,PDLA}.

\subsection{Calculation of $p_{\tau}(\Delta)$}

In this section we will derive the following approximate expression for
$p_{\tau}(\Delta)$
\begin{equation}
p_{\tau}(\Delta) \approx \left\{ \begin{array}{ll}
\frac{4}{3} \frac{\Gamma(\frac{2-\tau}{2})^2}
{\Gamma(2-\tau)} R^{1-\tau} \Delta^\tau  & \Delta \leq 1/2R \\
\frac{4}{3} \frac{\Gamma(\frac{2-\tau}{2})^2}
{\Gamma(2-\tau)} \frac{2^{\tau/2-3}R^{-\tau/2-2}}{\Delta^{3-3\tau /2}} &
\Delta > 1/2R  \ \ . \end{array}  \right.    \label{eqn:pcrude}
\end{equation}
This expression was used already in ref \cite{PDLA} but as it plays a central
role in this paper we present its derivation in some detail.
Complementary details may be found in refs \cite{PDsin,PDLA}.

We start from expression \EqRef{psos2} for $p_\tau (\Delta)$.
First we partition the SOS into subsets $\Omega_{\bf q}$ where
$\Omega_{\bf q}$ is the part of $\Omega$ for which the trajectory hits
disk ${\bf q}$. Then we smear the distribution, that is, we replace the delta
function with some extended distribution $\delta_\sigma$. The exact form
of this function is irrelevant,
the only thing we assume is that the width is big: $\sigma \gg 1$.
We will be interested in the behaviour of the leading zero of
$\hat{Z}_\tau (k)$ for small $\tau$.  This zero is then close to the origin and
it is evident that smearing of
$p_{\tau}(\Delta)$ will only have minor effect on this zero.

We now have
\begin{equation}
p_\tau(\Delta)=\sum_{\bf q} \int_{\Omega_{\bf q}} |\Lambda_s(x_s)|^{\tau}
\delta_\sigma (\Delta-\Delta_s(x_s)) dx_s  \ \ . \label{eqn:psoss}
\end{equation}
The large width $\sigma$ allow us to
move the smeared delta function
to the left of the integral sign
because the variation of
$\Delta_s(x_s) $
over $\Omega_{\bf q}$ is of the order $\sim R$ and we have $\sigma \gg 1 >R$.
\begin{equation}
p_\tau(\Delta)=\sum_{\bf q} \delta_\sigma (\Delta-q)
\int_{\Omega_{\bf q}} |\Lambda_s(x_s)|^{\tau}  dx_s \equiv
\sum_{\bf q} \delta_\sigma (\Delta-q) a_{\bf q}(\tau) \ \ .
\end{equation}
We have chosen the length of the lattice vector
$|{\bf q}| \equiv q$ as a mean value of
$\Delta_s(x_s)$. 
The local expansion factor is
$|\Lambda_s(x_s)|=2\Delta_s(x_s)/Rcos(\alpha)$.

It is easily shown that the phase space area
taken up by disk ${\bf q}$ is given by
the inequality
\begin{equation}
\mid \frac{q}{R}\sin (\phi - \theta_{\bf q}-\alpha)+\sin(\alpha)\mid < 1  \ \ ,
\label{eqn:olikhet}
\end{equation}
where $\theta_{\bf q}$ is the polar angle of the lattice vector ${\bf q}$.
Generally parts of this region are eclipsed by disks closer to the origin.
So in order to find $\Omega_{\bf q}$ one has to subtract these.
We will focus on the limit of small $R$ which gives the more easily handled
inequality
\begin{equation}
|\frac{q}{R} (\phi - \theta_{\bf q}-\alpha)+\sin(\alpha)| < 1  \ \ .
\label{eqn:appolikhet}
\end{equation}

Let us begin with
the limit of small $\Delta$.
In ref \cite{PDsin} it is shown that if
disk ${\bf q}$ lies within a certain radius: $q < 1/2R$, there are no
eclipsing disks in front of it, and expression \EqRef{appolikhet} may
be used directly, and the integral is easily calculated:
\begin{equation}
a_{\bf q}(\tau)  =
(\frac{2q}{R})^\tau \frac{1}{4\pi} \int_{\Omega_{\bf q}}
\cos^{\tau+1}(\alpha) d\alpha \; d\phi
\approx \frac{1}{4\pi} \frac{\Gamma(\frac{2-\tau}{2})^2}
{\Gamma(2-\tau)} (\frac{R}{q})^{1-\tau}  \   \  .
\label{eqn:aqfull}
\end{equation}
In order to find an approximate expression for $p_{\tau}(\Delta)$
we must know the
density of coprime lattice points, i.e. the average number
$d_c(r)dr$ of such points having
a distance between $r$ and $r+dr$ from the origin.
 In ref \cite{PDLA} this was, to leading order,
found to be $d_c(r)\approx
\frac{16\pi}{13}r$. This yields
\begin{equation}
p_{\tau}(\Delta)=\sum_{\bf q} \delta_\sigma (\Delta-q)
a_{\bf q}(\tau)=
\frac{16}{13} \frac{\Gamma(\frac{2-\tau}{2})^2}
{\Gamma(2-\tau)} R^{1-\tau} \Delta^\tau  \   \   .
\end{equation}
Obviously the disk radius has to be small for this to apply.

Next we consider the opposite limit $\Delta \gg 1/2R$.
For each (coprime) disk ${\bf q}$ fulfilling $q < 1/2R$ there are two
(or one, depending on symmetry)
transparent
corridor in the direction ${\bf q}$ \cite{PDsin,Bleh}, see fig 1b.
Far beyond this critical
radius the accessible disks will be those adjacent to the corridors. (They
still
have to coprime so they will lie on one side of the corridor only, see fig 1b).
We will discover that this will lead to a power law decay of
$p(\Delta)$. We will be interested in the particular power (as a function of
$\tau$) and not the prefactor. For that reason we perform our calculation in
the corridor having direction vector $(1,0)$ as all corridors provide the
same power. In this corridor the accessible disks are the ones being labeled
$(n,1)$. Disk  $(n,1)$ is shadowed by $(1,0)$ and $(n-1,1)$.
We need to evaluate the integral (cf eq \EqRef{aqfull})
\begin{equation}
j_n \equiv \int_{\Omega_{{\bf q}=(n,1)}}
\cos^{\tau+1}(\alpha) d\alpha \; d\phi  \ \ .
\label{eqn:jdef}
\end{equation}
It is more convenient to consider the sum
\begin{equation}
J_n \equiv \sum_{i=n}^{\infty} j_i=
\int_{\bigcup_{i=n}^{\infty} \Omega_{{\bf q}=(n,1)}}
\cos^{\tau+1}(\alpha) d\alpha \; d\phi \ \ ,
\label{eqn:Jdef}
\end{equation}
because this integral has support from the triangular region in fig 1c
(it is a triangle only in the limit $n\rightarrow \infty$ of course).

{}From eq \EqRef{appolikhet} one can  deduce that the base length
(in the $\phi$ direction) of this triangle
scales as $\sim 1/n$  and the height ( in the
$\alpha$ direction) as $1/\sqrt(n)$.
A short calculation now yields that
$J_n \sim 1/n^{2-\tau /2}$. Differentiation gives
$j_n \sim 1/n^{3-\tau /2}$. The fact that $l_{\bf q} \sim n$ together with eq.
\EqRef{aqfull} implies that the $a_{\bf q}$'s decay as $\sim 1/n^{3-3\tau /2}$.
The density of accessible disks is uniform in $\Delta$
(since they lie along
the corridors) so our final result is
\begin{equation}
\begin{array}{ll}
p(\Delta) \sim \frac{1}{\Delta^{3-3\tau /2}}  &
\Delta \gg \frac{1}{2R}  \label{eqn:farout} \ \ .
\end{array}
\end{equation}
This power law does not depend on the small $R$ limit.

The prefactor of eq \EqRef{farout}
is determined by demanding that $p_\tau (\Delta)$ is continous
at $\Delta =1/2R$.
In order to get correct normalization for $\tau =0$ we
multiply the entire  $p_\tau (\Delta)$ by $13/12$.
This approximation means a rather crude neglect of the of the transitional
behaviour at $\Delta \approx 1/2R$. In the following we must be very cautious
when using it since some results may be more sensible to our approximation than
other.

\subsection{Lyapunov exponents}

\begin{figure}
\epsffile{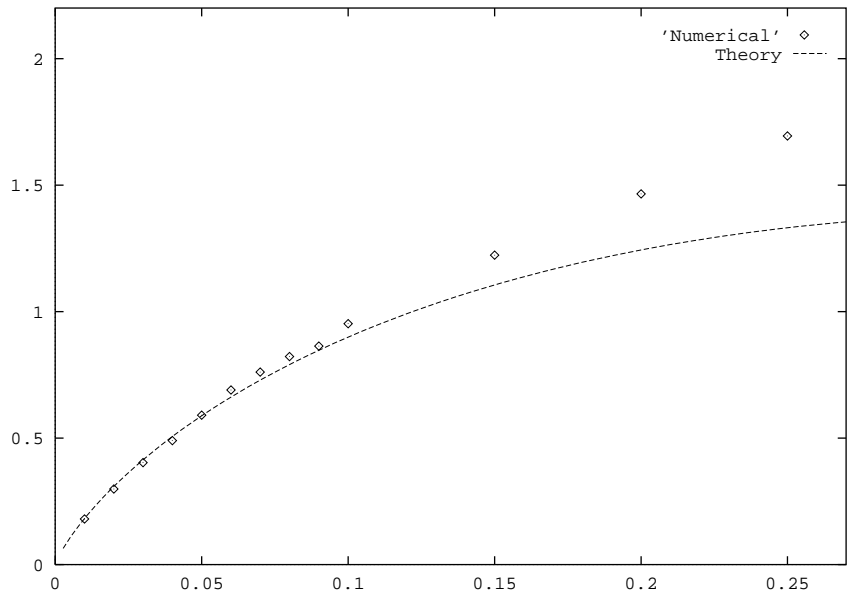}
\caption{Lyapunov exponent versus disk radius according to numerical simulation
and eq (32).}
\end{figure}

The approximate zeta function we are going to work with is obtained
from \EqRef{pcrude}
\begin{equation}
\begin{array}{ll}
\hat{Z}_{\tau}(k)\approx & 1-\int_0^{\infty} e^{-ik\Delta} p_\tau (\Delta)= \\
& 1-\frac{4}{3} \frac{\Gamma(\frac{2-\tau}{2})^2}
{\Gamma(2-\tau)} \frac{R^{-2\tau}}{2^{1+\tau}}(
z^{-1-\tau}\gamma (\tau+1,z)+
z^{2-3\tau /2} \Gamma(2+3\tau/2,z))  \ \ ,
\end{array}
\end{equation}
where  $z=\frac{ik}{2R}$.
The functions $\gamma(a,z)$ and $\Gamma(a,z)$ are incomplete gamma functions
\cite{AS}.
Expanding this to first order in $\tau$ gives
\begin{equation}
\begin{array}{lll}
Z=
& \left\{ z+\frac{z^2}{3}(\log z + \gamma -\frac{11}{6})\ldots \right\}+ & \\
& \left\{ (-\frac{7}{12}+\log (2R^2))+(\frac{11}{6}+\log (2R^2))z
\ldots  \right\}\tau & \ldots \end{array}
\label{eqn:ZLy}
\end{equation}
The derivation of this expansion is rather lengthy but the only step
that is slightly tricky is
the expansion of the incomplete gamma function
$\Gamma(a,z)$ near an integral power $a$. 
We have for convenience
thrown away all indices of the zeta function and
use strict equality sign but we must not forget that we work with an
approximation of an approximation of the exact zeta function.

We must now proceed with some care since the leading zero is not isolated,
a branch cut along the negative real $z$-axis reaches all the way up to it
(we have choosen the principal branch of the logarithm).

We are interested in the following derivative of the trace,
cf eq \EqRef{liatrace}
\begin{equation}
\frac{d}{d\tau} \; tr {\cal L}^t = \frac{d}{d\tau} \frac{1}{2\pi i}
\int e^{zt'}\frac{d}{dz} \log Z \; dz \mid_{\tau=0} =\frac{1}{2\pi i}
\int e^{zt'}\frac{d}{d\tau} \frac{d}{dz} \log Z \mid_{\tau=0} \; dz
  \ \ ,
\label{eqn:tracez}
\end{equation}
where we have differentiated inside the integral sign. We have now formulated
eq \EqRef{tracefour} in the rescaled and rotated $z-plane$ using rescaled time
$t'=2Rt$.
The function to be Fourier transformed is
\begin{equation}
\frac{d}{d\tau} \frac{d}{dz} \log Z \mid_{\tau=0} =
(\frac{7}{12}-\log (2R^2))\frac{1+\frac{z}{3}(2\log z + 2 \gamma -\frac{5}{6})
\ldots}{z+\frac{z^2}{3}(\log z +  \gamma -\frac{11}{6})^2 \ldots}=
(\frac{7}{12}-\log (2R^2))(\frac{1}{z^2}+\frac{1}{3z}\ldots )  \ \ ,
\label{eqn:f}
\end{equation}
where  we have kept only those terms yielding the leading term and the first
correction.  Collecting it all together yields
\begin{equation}
\lambda = \lim_{t\rightarrow 0} 2R (\frac{7}{12}-\log (2R^2) )
(1+\frac{1}{6Rt} \ldots )=2R (\frac{7}{12}-\log (2R^2) )  \  \ .
\end{equation}
The limiting value is due to the behaviour of the zero but the power law
correction are due to the fact that it sits on a branch point.
The particular size of the first correction is not very accurate, as we
will realize after reading section 3.4 and 3.5, but the important thing to bear
in mind is that there exist slowly decaying corrections
indicating slow convergence
of the Lyapunov exponent in numerical computations.

In fig 2 we compare numerical results on the Lyapunov exponent with
our expression $\lambda= 2R (\frac{7}{12}-\log (2R^2) )$.
The numerical values, from \cite{W},
are calculated for rather large disk radii, where we should
not expect much agreement a priori. Nevertheless our estimate is only
5\% wrong when $R=0.1$. The reason why the the numerical values exceed our
estimates for large $R$ is easily understood. This is because the disk faces
(in the unfolded system) come closer to each other so that taking the length
of the relevant lattice vector (as we did) overestimates the time of flight
between them.

The Lyapunov exponent of the corresponding Poincar\'{e} map with the disk
defining the SOS
is related to the Lyapunov exponent of the flow according to \cite{Abr}
$\lambda_{map}=\lambda <\Delta_s>\approx  \lambda /2R \approx
(\frac{7}{12}-\log (2R^2))$ where $<\Delta_s>$ is computed
by means of expression
\EqRef{pcrude}. We see that $\lambda \rightarrow -2\log(R)+7/12-\log 2$ when
$R \rightarrow 0$ which agrees with
the conjectured limit \cite{Oono,Bouch}:
$\lambda \rightarrow -2\log(R)+C+O(R)$.
Indeed we have analytically found an estimate of the constant $C\approx
7/12-\log 2 \approx -0.110$.
This is very close to the numerical value found by
\cite{Oono},
as far as we can extract it from fig 2 in ref \cite{Oono}).

In fig 3 we plot the generalized Lyapunov exponents for different disk radii.
Note that when $\tau >0$ the leading zero is indeed isolated
and we do not have to worry about the cut. The position of the zero is computed
numerically.
$\lambda(1)$ is the topological entropy which tend to a
finite limit when $R \rightarrow 0$  \cite{PDsin}, whereas $\lambda (0)
\rightarrow 0$ as $R \rightarrow 0$.
When $\tau < 0$ the branch cut itself will provide the leading behaviour of the
trace - a power law \cite{PDsin,PDLA}, and the generalized Lyapunov exponent
will
be zero. This means that
$\lambda(\tau)$ cannot be analytic at $\tau=0$.
This is referred to as a phase transition \cite{Beck,phase}.

\begin{figure}
\epsffile{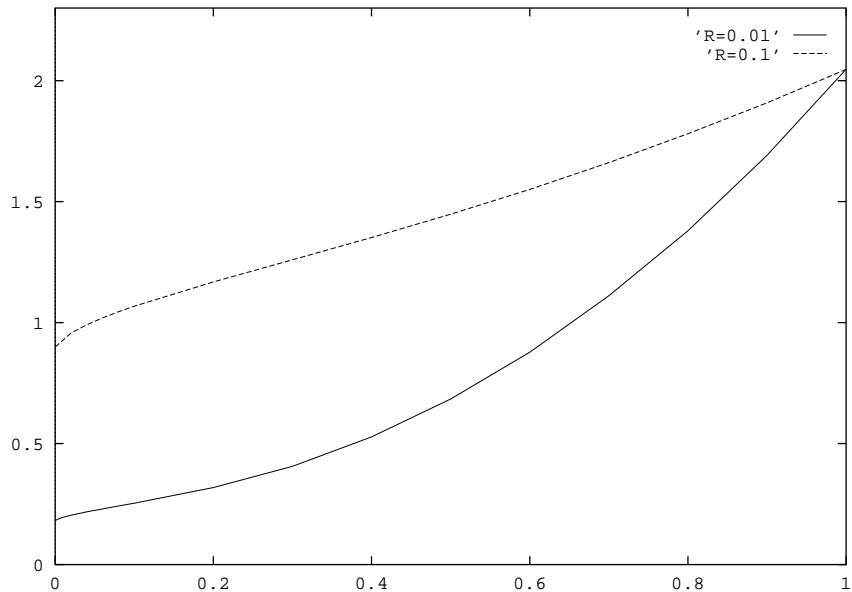}
\caption{Generalized Lyapunov exponents $\lambda(\tau)$ versus $\tau$
for two disk radii.}
\end{figure}

\subsection{A first calculation of the diffusion constant}

In order to study diffusion we calculate the generalized probability
distribution $p_w(\Delta)$ using appropriate weight \EqRef{diffw}.
However, we only keep the spatial components of $\beta$ yielding the
two dimensional vector ${\mbox{\boldmath $\beta$}}$.
As we study smeared $p_{\mbox{\boldmath $\beta$}}(\Delta)$ it suffices to
approximate the spatial part of $(\hat{f}^t(x)-x)$ with the lattice vector
{\bf q}.
So, we must now compute the generalized probability distribution
$p_{\mbox{\boldmath $\beta$}}(\Delta)$ a la section 3.1 but using the
weight
$\exp({\mbox{\boldmath $\beta$}} \cdot {\bf q})=\exp (\beta q cos(\phi_\beta -
\phi_q)$ where we have written ${\bf q}=q(\cos (\phi_q) , \sin (\phi_q))$ and
${\mbox{\boldmath $\beta$}}=\beta(\cos (\phi_\beta) , \sin (\phi_\beta))$.
We can use the result of section 3.1 to some extent,
since we realize, after
inspecting eq \EqRef{psos2}, that we can make the following factorization
\begin{equation}
p_{\mbox{\boldmath $\beta$}}(\Delta)=p_0(\Delta)
\frac{\int e^{\Delta \beta cos(\phi_q-\phi_\beta)} d\phi_q}{\int d\phi_q}
\end{equation}
To know the support of this integral
we need to know about the angular distribution of accessable disks for a
particular
value of $q=\Delta$. To this end we introduce
the following simplifying assumptions
\begin{enumerate}
\item  the coprime lattice points are distributed isotropically.
\item if $q > 1/2R$ we assume only four corridors with $\phi_q$ equal to one of
$0$, $\pi/2$, $\pi$ or $3\pi /2$.
\end{enumerate}
Assumption 2 means a rather brute neglect of the transitional
behaviour at $\Delta \approx 1/2R$.
We neglect the important fact
the number of corridors grows when $R \rightarrow 0$. We comment on this
in section 3.5. The important thing now is that we preserve the symmetry in
order to prevent net drift.

We first consider the case $q < 1/2R$. Using assumption 1 above gives
\begin{eqnarray}
p_{\mbox{\boldmath $\beta$}}(\Delta)=\frac{4}{3} R
\frac{1}{2\pi}\int_0^{2\pi}e^{\beta \Delta
\cos \phi }d\phi=\frac{4}{3} R I_0(\beta \Delta) = \frac{4}{3} R
(1+\frac{\beta^2 \Delta^2}{4} \ldots) & \Delta<\frac{1}{2R}  \  \ ,
\end{eqnarray}
where  $I_0$ is a modified Bessel function
In order to calculate the zeta function we need the Fourier transform of this
\begin{equation}
\int_0^{1/2R} p_{\mbox{\boldmath $\beta$}} (\Delta) e^{-ik\Delta} d\Delta=
\frac{2}{3}(1-z/2+z^2/6 \ldots) + \frac{\beta'^2}{6}(1/3-z/4+z^2/10 \ldots)  \
\ ,
\ldots
\end{equation}
where we have used rescaled variables $z=ik/2R$ and ${\mbox{\boldmath
$\beta$}}'={\mbox{\boldmath $\beta$}}/2R$.

Next we consider the limit $q > 1/2R$. We find, using the second assumption
above
\begin{eqnarray}
p_{\mbox{\boldmath $\beta$}}(\Delta)=\frac{2}{3(2R)^2} \frac{1}{\Delta^3}
(e^{\beta'_x \Delta}+
e^{-\beta'_x \Delta}+e^{\beta'_y \Delta}+e^{-\beta'_y \Delta}) &
\Delta > \frac{1}{2R}  \  \ .   \label{eqn:sture}
\end{eqnarray}
Fourier transforming this
gives
\begin{equation}
\int_{1/2R}^{\infty} p_{\mbox{\boldmath $\beta$}} (\Delta) e^{-ik\Delta}
d\Delta=
\frac{1}{6}(E_3(z-\beta'_x)+E_3(z+\beta'_x)+E_3(z-\beta'_y)+E_3(z+\beta'_y) )
\ \ .
\end{equation}
The resulting zeta function can now be expanded
\begin{equation}\begin{array}{rl}
Z_{\mbox{\boldmath $\beta$}}=1-
\int_{0}^{\infty} p_{\mbox{\boldmath $\beta$}} (\Delta) e^{-ik\Delta} d\Delta=
&
\frac{1}{6}(\gamma-\frac{11}{6})\beta'^2+z(1-\frac{1}{8}\beta'^2)+
z^2\frac{1}{3}(\gamma-\frac{11}{6})+ \\
\frac{1}{12} &
[ (z-\beta'_x)^2 \log (z-\beta'_x) + (z+\beta'_x)^2 \log (z+\beta'_x) +\\
& (z-\beta'_y)^2 \log (z-\beta'_y) + (z+\beta'_y)^2 \log (z+\beta'_y) ]\\
& \ldots
\end{array} \label{eqn:Zbeta} \end{equation}
According to sec 2.2, we are interested in the following quantity
\begin{equation}\begin{array}{rl}
(\frac{\partial^2}{\partial \beta_x^2}+\frac{\partial^2}{\partial \beta_y^2})
tr {\cal L} \mid_{{\bf \beta'}=0} = &
(\frac{\partial^2}{\partial \beta_x^2}+\frac{\partial^2}{\partial \beta_y^2})
\frac{1}{2\pi i}
\int e^{zt'}\frac{d}{dz} \log Z \; dz \mid_{\beta=0} \\
 = &
\frac{1}{(2R)^2}\frac{1}{2\pi i}
\int e^{zt'}
(\frac{\partial^2}{\partial \beta_x^{'2}}+\frac{\partial^2}{\partial
\beta_y^{'2}})  \frac{d}{dz} \log Z \mid_{\beta=0} \; dz
  \ \ .
\label{eqn:tracebeta}
\end{array}\end{equation}
We now proceed in  the same way as in sec 3.2. We want to determine the
asymptotic behaviour of the fourier transform of
\begin{equation}
(\frac{\partial^2}{\partial \beta_x^{'2}}+\frac{\partial^2}{\partial
\beta_y^{'2}})  \frac{d}{dz} \log Z \mid_{{\mbox{\boldmath $\beta$}}=0} \sim
\frac{2}{3z^2}(\frac{4}{3}-\log z -\gamma) \ \ .
\label{eqn:g}
\end{equation}
To this end we now need the following integral
\begin{equation}
\frac{1}{2\pi i}\int \frac{\log z}{z^2} e^{zt'}d z=
t' (1-\gamma-log t')  \ \ .
\end{equation}
In deriving this it is convenient to use contour integration and let the
contour
encircle the negative real $z$ axis.
We thus find the following diverging diffusion constant
($D=\lim_{t \rightarrow \infty} D(t)$)
\begin{equation}
D(t)=\frac{1}{2t}
(\frac{\partial^2}{\partial \beta_x^2}+\frac{\partial^2}{\partial \beta_y^2})
tr {\cal L}^t \mid_{{\bf \beta'}=0}=
\frac{1}{18R} (1+3log(2Rt)) \label{eqn:Dt} \  \ .
\end{equation}
This logarithmic divergence of the diffusion constant agrees with
ref \cite{Bleh} but the prefactor is not correct.
The computation will be refined in the section 3.5.
The important thing to learn from this section is that the Laplacian
exposes the logarithm in the series expansion of the zeta function. So we
now know where to focus our attention, namely at the tail of
$p_{\mbox{\boldmath $\beta$}}(\Delta)$.

\subsection{A closer look at the tail}

At the end of section 3.1 we aimed at finding a good approximation
for $p_\tau(\Delta)$ for the whole range $1/2R< \Delta < \infty$.
In this section we will make a more careful investigation of the
$\Delta \rightarrow \infty$
tail of $p_\tau(\Delta)$.
We restrict ourselves to the case $\tau=0$ and will compute the limit
$\lim_{\Delta \rightarrow \infty} \Delta^3 p_0(\Delta)$ for small disk radii.

Let us look at the corridor with direction vector ${\bf q}=(n_x,n_y)$ where
$n_x$ and $n_y$ are coprime.
Suppose for
a moment that ${\bf q}$ lies in the first octant so that $n_x$ and $n_y$
positive
and $n_x\geq n_y$.
The accessible disks in this corridor are the ones labeled
${\bf q}'+n{\bf q}$ and ${\bf q}''+n{\bf q}$
where ${\bf q}'$ and ${\bf q}''$ are the
predecessor and successor in the Farey sequence of order $n_x$ \cite{PDsin}.
A calculation
analogous to the one at the end of section 3.1 now gives the following
expression for $a_q(0)$ (as defined in eq \EqRef{aqfull})
\begin{equation}
a_{{\bf q}'+n{\bf q}}=\frac{1}{2\pi}\frac{2qR+\frac{1}{2qR}-2}{q^2n^3}+O(1/n^4)
\end{equation}
and the same holds for the sequence $q''+nq$.
The contribution to $p_0(\Delta)$ is
\begin{equation}
\sum_n a_{q'+nq} \delta(\Delta-|q'+nq|)=\frac{1}{\Delta^3}\frac{1}{2\pi}
(2qR+\frac{1}{2qR}-2)
\end{equation}
To compute the tail of $p_0(\Delta)$ we need to sum over all coprime
${\bf q}$ such that $q<1/2R$. We restrict the summation to the first
quadrant ${\bf q} \in S$
\begin{equation}
S=\{ {\bf q}=(n_x,n_y) | n_y>0; \; n_x \geq 0; \; (n_x,n_y)=0; \;
\sqrt{n_x^2+n_y^2}<1/2R  \}
\end{equation}
and multiply the result by four, to account for all four quadrants, and then by
two, to account for both ${\bf q}'$ and ${\bf q}''$ defined above.
The result is
\begin{equation}
p_0(\Delta) \sim  \frac{4}{\pi\Delta^3} \sum_{{\bf q} \in S}
(2qR+\frac{1}{2qR}-2) +O(1/\Delta^4)
\end{equation}
The nontrivial sum above is treated by Bleher \cite{Bleh} and we insert the
small $R$ limit of the sum and obtain
\begin{equation}
p_0(\Delta) \rightarrow \frac{1}{\Delta^3 2 \pi^2 R^2}+O(1/\Delta^4)
\label{eqn:p0as}
\end{equation}

Suppose that we expand the unweighted zeta function $Z_0$ into the more
general series
\begin{equation}
Z_0=\sum_{i=1}^{\infty} a_i z^i +
\sum_{i=2}^{\infty} b_i z^i\log z   \label{eqn:Psi}
\end{equation}
For the approximation of $p(\Delta)$ worked out in section 3.1 we have
$b_i=0$ for $i\geq 3$.
The correct value of $b_2$ may be computed from eq \EqRef{p0as} (using e.g.
standard expansions of exponential integrals) and is
found to be $b_2=1/\pi^2$
which differs considerably from the result in
section 3.1.
The reason for this error is the absence of a smooth transition
at $\Delta=1/2R$ in eq \EqRef{pcrude}. The crossover behaviour is slow
as indicated from the $O(1/\Delta^4)$ term above.
Generally we expect the asymptotics of the tail
to look like $p(\Delta) \sim \sum_{n\geq 3} c_n/\Delta^n$ so that there now
may exist nonzero $b_i$ of any order.

\subsection{A second calculation of the diffusion constant}

One could suspect that the coarse assumptions,
especially assumption 2 in section
3.3, induce an error in the prefactor of super diffusion.
We will now demonstrate that the prefactor is robust against refinements
of the this assumption. A better estimate of $p_\beta(\Delta)$ is obtained
by replacing $exp(\beta'_x \Delta)=exp(\beta' \Delta \cos (\phi_\beta))$
in eq \EqRef{sture} with
the averaged quantity
\begin{equation}
\frac{1}{2c/\Delta}\int_{-c/\Delta}^{c/\Delta}
e^{\beta' \Delta\cos (\phi_\beta-\phi_q)}d\phi_q
\end{equation}
The integration range decreases like $1/\Delta$ since
the width of the corridor is constant.
For sufficiently large $\Delta$ we can take
$\beta' \cos (\phi_\beta-\phi)\approx \beta'_x+\beta'_y \phi$
and find that the integral above is
\begin{equation}
e^{\beta'_x \Delta} (1+O({\beta'_y}^2)
\end{equation}
The correction is $O(\beta'^2)$
and has no effect on $D(t)$, as we anticipated.

We also reduced the number of corridors to four, irrespective of the disk
radius. As long as we are not interested in the angular properties this
is easily justified by studying how the laplacian acts on \EqRef{Zbeta}.

We are led to the conclusion that
the prefactor of super diffusion depends solely on the coefficient
$b_2$ of the unweighted zeta function \EqRef{Psi}
according to $D(t)=\frac{b_2}{2R}\log t$.
Inserting the correct value of $b_2$ from the previous section we arrive at
\begin{equation}
D(t)=\frac{1}{2R\pi^2}\log t
\end{equation}
which is indeed the exact result \cite{Bleh}.

Again there are slowly decaying corrections to $D(t)$ and it is not surprising
that the numerical detection of this diffusion law has eluded serious attempts
\cite{Gala}.

\section{Concluding remarks}

The main advantage of the approximation outlined in this paper is its apparent
simplicity. It is far more easy to apply than a proper cycle expansion would
be, taking all the (infinite number of) pruning rules into account, and it
gives a very good description of the leading zero.
This zero and its vicinity, was our
only interest in this paper and we could work out a very coarse description
of $p(\Delta)$.

In ref. \cite{PDsin} we demonstrated that refinement of the expression for
$p(\Delta)$ down to a scale $\sim R$ also accounted for a good description of
the trace down to this scale. This means that some nonleading zeros
may also be extracted within the BER approximation. However, in order to
compute
higher order spectra, like semiclassical spectra,
one needs to take the correlation between laminar intervals
into account.
The way to do this would perhaps be to combine
the knowledge of the asymptotics on the periodic orbits, as obtained from the
present approximation, with a set of explicitly computed periodic orbits.

The major drawback of the BER approximation is the lack of error term; so
far we have no bounds on the errors in our computations. It is of course highly
desirable to work out such bounds which would put the present considerations on
mathematically firm ground.

\section*{Acknowledgements}

I would like to thank Predrag Cvitanovi\'{c} who suggested anomalous diffusion
to be a suitable subject of our approximation scheme.
The approach in section 2.1 follows a lecture by Hans Henrik Rugh
rather closely.
This work was supported by the Swedish Natural Science
Research Council (NFR) under contract no. F-FU 06420-303.

\newcommand{\PR}[1]{{Phys.\ Rep.}\/ {\bf #1}}
\newcommand{\PRL}[1]{{Phys.\ Rev.\ Lett.}\/ {\bf #1}}
\newcommand{\PRA}[1]{{Phys.\ Rev.\ A}\/ {\bf #1}}
\newcommand{\PRD}[1]{{Phys.\ Rev.\ D}\/ {\bf #1}}
\newcommand{\PRE}[1]{{Phys.\ Rev.\ E}\/ {\bf #1}}
\newcommand{\JPA}[1]{{J.\ Phys.\ A}\/ {\bf #1}}
\newcommand{\JPB}[1]{{J.\ Phys.\ B}\/ {\bf #1}}
\newcommand{\JCP}[1]{{J.\ Chem.\ Phys.}\/ {\bf #1}}
\newcommand{\JPC}[1]{{J.\ Phys.\ Chem.}\/ {\bf #1}}
\newcommand{\JMP}[1]{{J.\ Math.\ Phys.}\/ {\bf #1}}
\newcommand{\JSP}[1]{{J.\ Stat.\ Phys.}\/ {\bf #1}}
\newcommand{\AP}[1]{{Ann.\ Phys.}\/ {\bf #1}}
\newcommand{\PLB}[1]{{Phys.\ Lett.\ B}\/ {\bf #1}}
\newcommand{\PLA}[1]{{Phys.\ Lett.\ A}\/ {\bf #1}}
\newcommand{\PD}[1]{{Physica D}\/ {\bf #1}}
\newcommand{\NPB}[1]{{Nucl.\ Phys.\ B}\/ {\bf #1}}
\newcommand{\INCB}[1]{{Il Nuov.\ Cim.\ B}\/ {\bf #1}}
\newcommand{\JETP}[1]{{Sov.\ Phys.\ JETP}\/ {\bf #1}}
\newcommand{\JETPL}[1]{{JETP Lett.\ }\/ {\bf #1}}
\newcommand{\RMS}[1]{{Russ.\ Math.\ Surv.}\/ {\bf #1}}
\newcommand{\USSR}[1]{{Math.\ USSR.\ Sb.}\/ {\bf #1}}
\newcommand{\PST}[1]{{Phys.\ Scripta T}\/ {\bf #1}}
\newcommand{\CM}[1]{{Cont.\ Math.}\/ {\bf #1}}
\newcommand{\JMPA}[1]{{J.\ Math.\ Pure Appl.}\/ {\bf #1}}
\newcommand{\CMP}[1]{{Comm.\ Math.\ Phys.}\/ {\bf #1}}
\newcommand{\PRS}[1]{{Proc.\ R.\ Soc. Lond.\ A}\/ {\bf #1}}

\newpage

\section*{Appendix}

In this appendix we will verify equation \EqRef{start}
$<w>=\lim_{t \rightarrow \infty} tr {\cal L}_w^t$.

The invariant density may be expressed in terms of periodic orbits
as $\mu(dx)=\lim_{t \rightarrow \infty}\mu_t(dx)$ where
\begin{equation}
\mu_t (dx) = dx \sum_{p \in p.p.o.}
\sum_{n=1}^{\infty}  \frac{\delta(t-nT_p)}{\Abs{det(1-M_p^n)}}
\int_{x' \in p} \delta
(x-x') \frac{dx'}{v}
\end{equation}
where $v$ is the speed at $x$.

The invariant density is a eigenfunction of the
evolution operator ($w=1$) corresponding to the leading eigenvalue
(equal to one).
This requirement is satisfied by $\mu_t(dx)$ for any $t$.
However, $\mu_t(dx)$ is only a smooth function in the limit
${t \rightarrow \infty}$ (we refrain from showing this).
It follows from the boundedness that the density is normalized
$\int \mu_t(dx) \rightarrow 1$ as $t \rightarrow \infty$ \cite{Ozo,flows}.
The expectation value
can now be expressed in terms of periodic orbits as
\begin{eqnarray}
<w>=\int w(x,t) \mu(dx)=\\
\lim_{t \rightarrow \infty}
\sum_{p \in p.p.o.} T_p \sum_{n=1}^{\infty} w_p^n
\frac{\delta(t-nT_p)}{\Abs{det(1-M_p^n)}}
\end{eqnarray}
Comparing this expression with eq \EqRef{tracedef} immediately
verifies eq \EqRef{start}.

\end{document}